\documentclass[aps,prl,reprint,superscriptaddress]{revtex4-2}
\usepackage{graphicx}
\usepackage{amsmath}
\usepackage{mathcomp}
\usepackage{textcomp}
\usepackage{siunitx}
\usepackage[english]{babel}

\usepackage{placeins} %for FloatBarrier
\usepackage{braket}

\begin{document}

% Use the \preprint command to place your local institutional report
% number in the upper righthand corner of the title page in preprint mode.
% Multiple \preprint commands are allowed.
% Use the 'preprintnumbers' class option to override journal defaults
% to display numbers if necessary
%\preprint{}

%Title of paper
\title{On-chip single-photon subtraction by individual silicon vacancy centers in a laser-written diamond waveguide}

% repeat the \author .. \affiliation  etc. as needed
% \email, \thanks, \homepage, \altaffiliation all apply to the current
% author. Explanatory text should go in the []'s, actual e-mail
% address or url should go in the {}'s for \email and \homepage.
% Please use the appropriate macro foreach each type of information

% \affiliation command applies to all authors since the last
% \affiliation command. The \affiliation command should follow the
% other information
% \affiliation can be followed by \email, \homepage, \thanks as well.
\author{Michael K. Koch}
\email[M.K.K. and M.H. contributed equally to this work.]{}
\affiliation{Institute for Quantum Optics, Ulm University, D-89081 Ulm, Germany}
\affiliation{Center for Integrated Quantum Science and Technology (IQst), Ulm University, D-89081 Ulm, Germany}

\author{Michael Hoese}
\email[M.K.K. and M.H. contributed equally to this work.]{}
\affiliation{Institute for Quantum Optics, Ulm University, D-89081 Ulm, Germany}
%
%\author{Prithvi Reddy}
%\affiliation{Laser Physics Centre, Research School of Physics and Engineering, %Australian National University, Canberra, ACT 0200, Australia}

\author{Vibhav Bharadwaj}
\affiliation{Institute for Photonics and Nanotechnologies (IFN) - CNR, Piazza Leonardo da Vinci, 32, Milano, 20133, Italy}

\author{Johannes Lang}
\affiliation{Institute for Quantum Optics, Ulm University, D-89081 Ulm, Germany}
\affiliation{Diatope GmbH, D-88444 Ummendorf, Germany}

\author{John P. Hadden}
\affiliation{School of Physics and Astronomy, Cardiff University, Cardiff CF24 3AA, United Kingdom}

%\author{Reina Yoshizaki}
%\affiliation{Department of Mechanical Engineering, School of Engineering, The University of Tokyo, Tokyo, 113-8656, Japan}

%\author{Argyro N. Giakoumaki}
%\affiliation{Institute for Photonics and Nanotechnologies (IFN) - CNR, Piazza Leonardo da Vinci, 32, Milano, 20133, Italy}

\author{Roberta Ramponi}
\affiliation{Institute for Photonics and Nanotechnologies (IFN) - CNR, Piazza Leonardo da Vinci, 32, Milano, 20133, Italy}

\author{Fedor Jelezko}
\affiliation{Institute for Quantum Optics, Ulm University, D-89081 Ulm, Germany}
\affiliation{Center for Integrated Quantum Science and Technology (IQst), Ulm University, D-89081 Ulm, Germany}

\author{Shane M. Eaton}
\affiliation{Institute for Photonics and Nanotechnologies (IFN) - CNR, Piazza Leonardo da Vinci, 32, Milano, 20133, Italy}

\author{Alexander Kubanek}
\email[Corresponding author: ]{alexander.kubanek@uni-ulm.de}
\affiliation{Institute for Quantum Optics, Ulm University, D-89081 Ulm, Germany}
\affiliation{Center for Integrated Quantum Science and Technology (IQst), Ulm University, D-89081 Ulm, Germany}

%Collaboration name if desired (requires use of superscriptaddress
%option in \documentclass). \noaffiliation is required (may also be
%used with the \author command).
%\collaboration can be followed by \email, \homepage, \thanks as well.
%\collaboration{}
%\noaffiliation

\date{\today}

\begin{abstract}
Modifying light fields at single-photon level is a key challenge for upcoming quantum technologies and can be realized in a scalable manner through integrated quantum photonics.
Laser-written diamond photonics offers three-dimensional fabrication capabilities and large mode-field diameters matched to fiber optic technology, though limiting the cooperativity at the single-emitter level.
To realize large cooperativities, we combine excitation of single shallow-implanted silicon vacancy centers via large numerical aperture optics with detection assisted by laser-written type-II waveguides.
We demonstrate single-emitter extinction measurements with a cooperativity of 0.153 and a beta factor of 13$\si{\percent}$ yielding 15.3$\si{\percent}$ as lower bound for the quantum efficiency of a single emitter.
The transmission of resonant photons reveals single-photon subtraction from a quasi-coherent field resulting in super-Poissonian light statistics.
Our architecture enables single quantum level light field engineering in an integrated design which can be fabricated in three dimensions and with a natural connectivity to optical fiber arrays.

\end{abstract}

% insert suggested keywords - APS authors don't need to do this
%\keywords{}

%\maketitle must follow title, authors, abstract, and keywords
\maketitle

% body of paper here - Use proper section commands
% References should be done using the \cite, \ref, and \label commands
\section{Introduction}

%,offer a versatile alternative platform that allows three-dimensional writing capabilities

Color centers in diamond are fast becoming an established platform for solid-state based quantum communication and quantum sensing
\cite{Atature2018}. These color centers offer optically addressable electronic spin qubits \cite{Awschalom2018}, access to nuclear spins \cite{Neumann2010, Metsch2019}, which can serve as quantum memory \cite{Maurer2012, Bradley2019}, and can be created in deterministic approaches \cite{Evans2016, Chen2017, Schroder2017}. Their potential for the realization of quantum networks is highlighted by the recent realization of a three-node quantum network with nitrogen vacancy (NV$^-$) centers \cite{Pompili2021} and a quantum memory node with a silicon vacancy (SiV$^-$) center \cite{Bhaskar2020}. Integration into photonic platforms with efficient coupling and access to a quantum optical non-linearity is a key challenge to facilitate scalability and robust operation for beyond-demonstrator implementation \cite{Ruf2021}. 
Diamond-based nanophotonics has reached the highest level of complexity \cite{Hausmann2012, Riedrich2014, Zhang2018}, including new design paths such as inverse engineering \cite{Dory2019}. Its performance is ultimately limited by the demands of complexity and, in particular, on the required purity of the fabrication process. Working with photonics with larger mode-field diameters and therefore larger distance of the color center to the diamond surface would help to reduce the requirements on the fabrication process but inevitably leads to a reduced light-matter interaction strength prohibiting applications in the quantum regime.
Another approach is hybrid quantum photonics, which offers the advantage of separate optimization of the photonics and the quantum emitter. However, the hybrid approach relies on elaborate manual functionalization \cite{Schrinner2020, Wan2020, Fehler2021}. 

Laser-written photonic structures in diamond, based on type-II waveguides, offer a scalable pathway \cite{Wan2020}, rapid fabrication, and enables incorporation of color centers \cite{Sotillo2016}. The laser-written waveguides can be adapted to a broad wavelength range and extended to more complex photonic structures such as beamsplitters \cite{Courvoisier2016}, Bragg-mirrors \cite{Bharadwaj2017} and 3D structures \cite{Courvoisier2016, Kononenko2011}. To functionalize the platform, NV$^-$ centers have been created on demand by laser iradiation \cite{Hadden2018} or by shallow ion implantation into the front facet \cite{Hoese2021}. 
However, the intrinsic cooperativity of laser-written photonics is still low owing to the relatively large mode-field diameter, therefore prohibiting non-linear effects on the single photon single emitter level. 

Here, we overcome this limitation by exciting individual emitters with a high numerical aperture (NA) while the detection is realized with laser-written waveguides in diamond. This combination facilitates a high cooperativity and, at the same time, efficient light field engineering on chip and in three-dimensional waveguide arrays. We begin by functionalizing the waveguides with SiV$^-$ centers, one of the group-IV color centers in diamond, with a high coherent photon flux due to its high Debye-Waller factor \cite{Neu2011, Dietrich2014}, an exceptional spectral stability due to its inversion symmetry \cite{Rogers2014b, Sipahigil2014} and coherent optical controllability \cite{Rogers2014c, Pingault2017, Becker2018}. Due to the strong interaction with light, single SiV$^-$ centers can significantly modify weak light fields. We first perform extinction measurements to quantify the light-matter interaction on the single emitter level. We then use the coupled system to subtract individual photons from a quasi-coherent light field in order to create a light field with super-Poissonian light statistics. Our results open the door for scalable, on-chip light field engineering on the single photon level.

\section{Experimental Setup}
\begin{figure}[]
\includegraphics[scale=1]{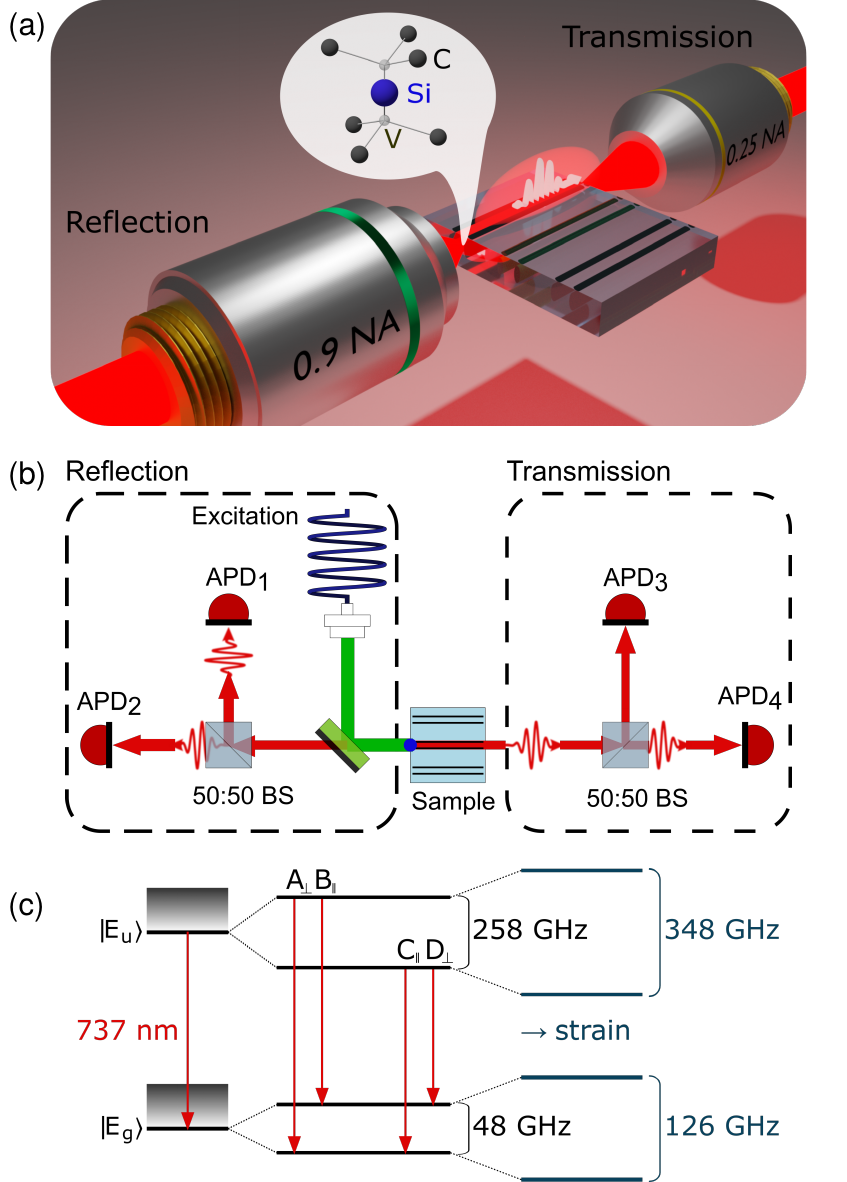}
\caption{\label{Fig:setup-overview} Experimental outline. (a) Schematic depiction of single photon coupling in the diamond waveguide. Here, the guidance of the excitation laser and the photon emission of a single shallow implanted SiV$^-$ center is depicted. (b) Setup sketch, where the emitter is excited in the reflection path. Each detection path, in reflection and transmission, is equipped with two single photon detectors in HBT configuration. (c) Energy level structure of the utilized SiV$^-$ defect center. Here, the ground and excited state splitting is shown for the zero strain case and the strain induced splittings that we observe.}
\end{figure}

We use laser-written type-II waveguides in an electronic grade diamond slab of dimensions $2\si{\milli\meter} \times 2\si{\milli\meter} \times 0.3\si{\milli\meter}$. Each waveguide consists of two walls, indicated by the black lines in the sketch of the experimental setup in fig. \ref{Fig:setup-overview} (a) and (b). These walls are created by femtosecond laser-writing \cite{Sotillo2016, Eaton2019, Bharadwaj2019} and are optimized for a transmission wavelength of $738\si{\nano\meter}$, corresponding to the zero phonon line (ZPL) wavelength of the SiV$^-$ center. Similar to previous experiments with NV$^-$ centers \cite{Hoese2021}, the defect centers are created by shallow ion implantation into the front facet of the waveguides (see fig. \ref{Fig:setup-overview} (a)) \cite{Lang2020}. In order to obtain spectrally stable emitters we anneal the sample at $1000\si{\celsius}$ after implantation. For more information on the implantation and annealing procedure, see supplementary information.
%\cite{Lang2020}

The presence of the waveguide walls induces strain on the SiV$^-$ centers, which not only shifts the wavelength of the ZPL but also increases its ground state and excited state splitting \cite{Meesala2018}. The energy level scheme of the SiV$^-$ center with its characteristic four-line structure (transition A to D) is shown in fig. \ref{Fig:setup-overview} (c). The strain-induced shifts that we observe are also shown in this level scheme. We observe a ground and excited state splitting of $126 \si{\giga\hertz}$ and $348 \si{\giga\hertz}$, respectively.

The experimental setup, illustrated in fig. \ref{Fig:setup-overview} (b), can be divided into two parts, the reflection and transmission path. A 0.25 NA objective aligned with one of the waveguides is used in the transmission path to collect the SiV$^-$ emission coupled into the waveguide. We have confocal access via the reflection channel (0.9 NA), where the facet of the sample can be scanned using off resonant $532\si{\nano\meter}$ laser excitation or on-resonance with a tunable laser excitation source. Both transmission and reflection paths can be used in Hanbury Brown-Twiss (HBT) configuration to perform correlation measurements. All measurements are conducted at $5\si{\kelvin}$ in a flow cryostat.

\section{Device Characterization}
\begin{figure*}[]
\includegraphics[scale=1.0]{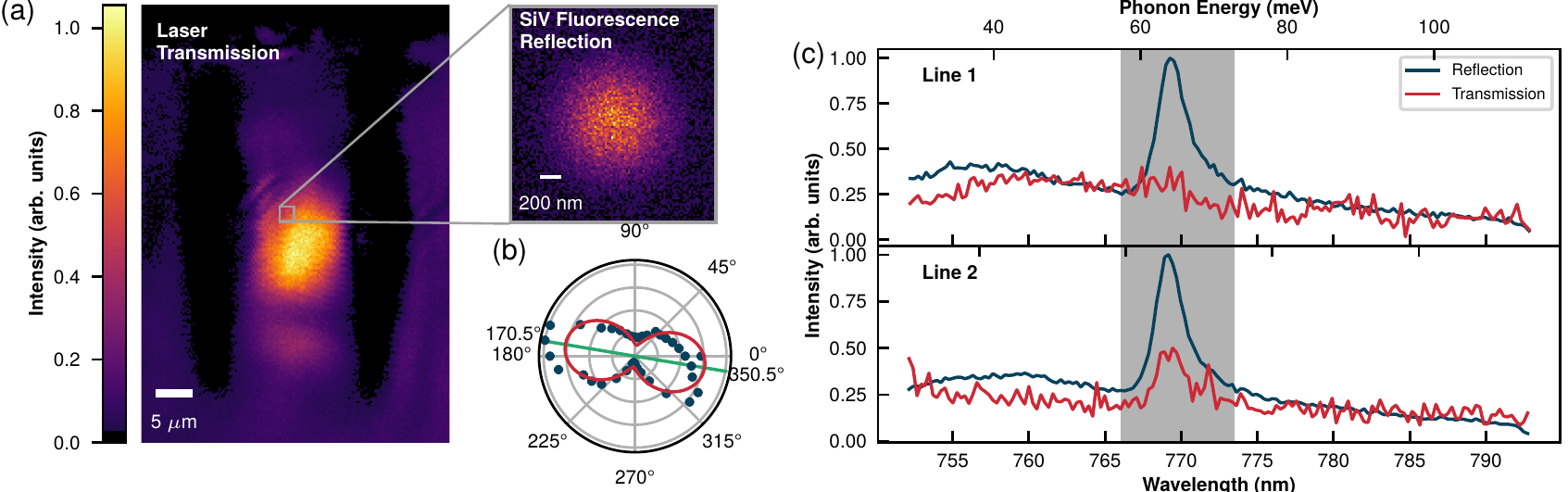}
\caption{\label{Fig:sample-characterization}Device characterization. (a) Confocal scan (detected in transmission) of the the waveguide, to map the waveguide mode. The laser wavelength is set to $738\si{\nano\meter}$ and the polarization is set to the polarization of the waveguide. The inset shows a confocal scan of the SiV center in reflection at resonance ($\lambda = 739.338\si{\nano\meter}$) and the markings reveal the position of the defect center within the waveguide. (b) Polarization dependence of the waveguide, measured in the same configuration as in part (a), but the laser is placed at the waveguide mode maximum. (c) Spectrum of the PSB for the PLE measurement for two different transitions, Line 1 at $738.404\si{\nano\meter}$ and Line 2 at $738.327\si{\nano\meter}$, in reflection and transmission for comparison. The grey shaded area denotes the local mode at $64\si{\milli\electronvolt}$.}
\end{figure*}

First, we map the waveguide mode by scanning the high-NA objective for excitation in the reflection path with a $738\si{\nano\meter}$ laser and collect the coupled laser light via the transmission channel. The mode profile is depicted in fig. \ref{Fig:sample-characterization} (a), where the black vertical bars correspond to the walls of the waveguide. Next we excite a single defect center resonantly using $739.338\si{\nano\meter}$ laser light. The position of the defect center is indicated within the waveguide mode, together with a high resolution confocal scan performed in reflection. The polarization pattern of the waveguide (see fig. \ref{Fig:sample-characterization} (b)), measured at the waveguide mode maximum, is aligned at $\left( 170.5 \pm 2.4 \right)\si{\degree}$.
To test the polarization selectivity of the waveguide, we resonantly excite two different SiV$^-$ defect centers (Line 1 and 2) near the transmission maximum of the waveguide and detect their respective phonon side bands (PSBs) in both detection channels. The detection in the reflection channel gives us the fluorescence signal independent of the waveguide transmission properties and allows us to compare it to the signal guided through the waveguide (transmission channel), which is polarization selected according to the waveguide properties. Accordingly, the polarization supported by the waveguide transmission is more pronounced in the transmission signal. Two distinct features are the broad and the local mode with energy differences of $41\si{\milli\electronvolt}$ and $64\si{\milli\electronvolt}$ from the ZPL, respectively \cite{Rogers2014}. The measured PSBs of both Line 1 and 2, are shown in fig. \ref{Fig:sample-characterization} (c). In reflection, the PSB appears almost identical for both lines. In transmission, the local mode of Line 1 is much less visible compared to the $41\si{\milli\electronvolt}$ mode and the reflection signal. The local mode of Line 2, on the other hand, is transmitted much better, which means that Line 1 is polarized differently than Line 2.
The PL spectrum of the SiV$^-$ center is shown in fig. \ref{Fig:ResonantTransmission} (a). The labels A to D denote the transitions illustrated in fig. \ref{Fig:setup-overview} (c). For resonant excitation we use transition D. The polarization pattern for the emission dipole of transition D and the local mode is plotted in the inset of fig. \ref{Fig:ResonantTransmission} (a). Note that the polarization pattern of the local mode matches the waveguide polarization almost perfectly, while due to strain the polarization of transition D is tilted by $45\si{\degree}$. For comparison and later analysis of the emitter coupling to the waveguide, we perform a PLE scan of transition D in reflection and transmission with a long scan time of $100 \si{\second}$, shown in fig. \ref{Fig:ResonantTransmission} (b).
 
%738.328nm and 739.338nm
%peaks in PL spec @ 738.345nm, 738.812nm, 739.046nm, 739.451nm, 739.679nm

\section{Extinction Measurement}

\begin{figure*}[]
\includegraphics[scale=1.0]{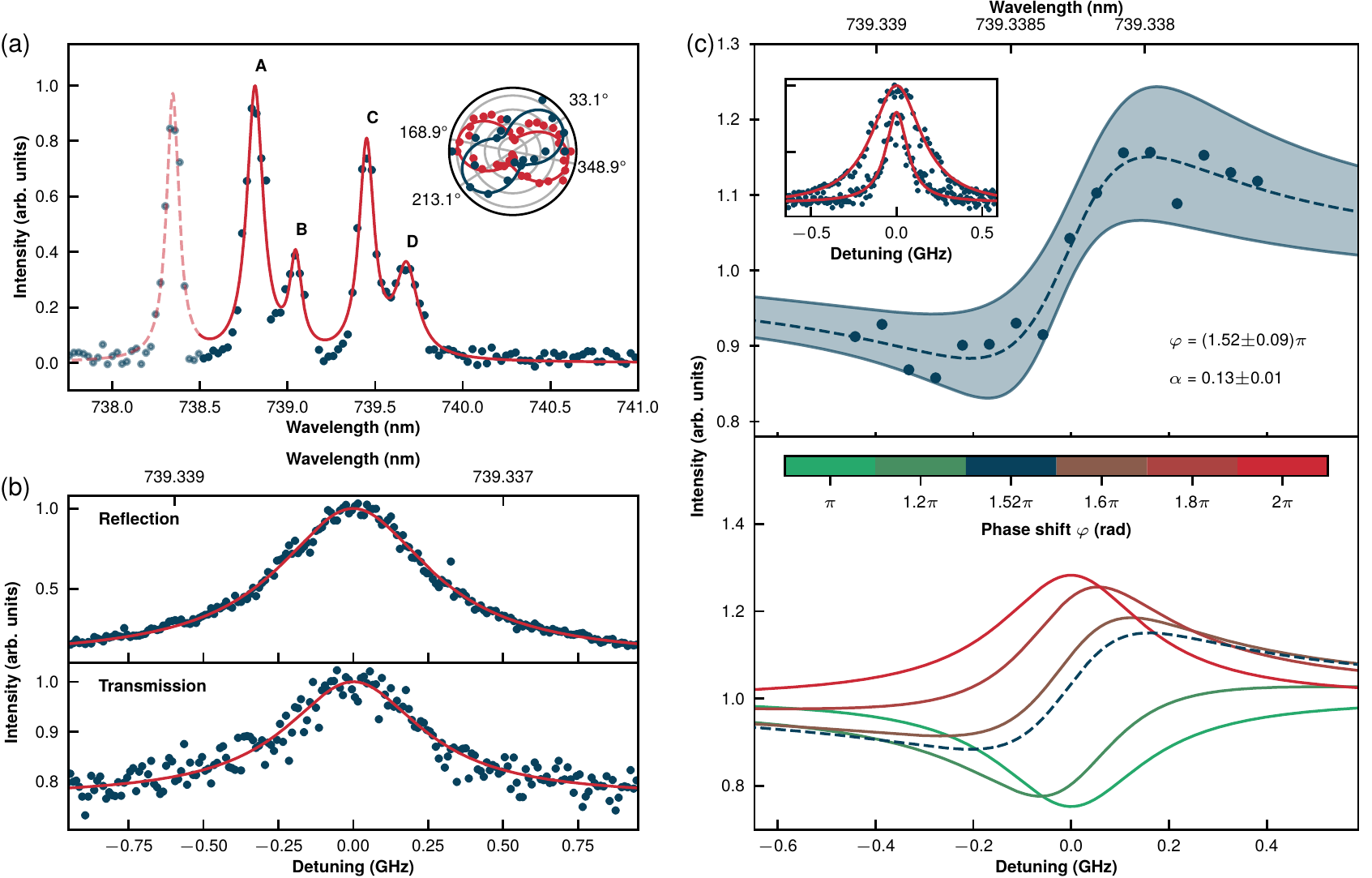}
\caption{\label{Fig:ResonantTransmission} Resonant transmission 
(a) PL spectrum of the emitter (excited with $532\si{\nano\meter}$). The spectrum is background corrected and normalized to the maximum value of the corresponding fit function. The first peak in the spectrum, indicated by the dashed line, belongs to a second emitter within the confocal spot of the excitation laser. The inset polar plot shows the emission dipole of transition D (blue curve, off-resonantly excited) and the local mode (red curve, resonantly excited at $739.338\si{\nano\meter}$). 
(b) PLE scan of the emitter with a center wavelength of $739.338\si{\nano\meter}$ detected in both, the reflection and transmission channel. 
(c) Resonant transmission spectrum. The upper panel shows the measured transmission spectrum of the defect center waveguide system, where for each data point a 30 second count trace was recorded. Here the dashed line is the resulting fit function fitted with eq. \eqref{eq:Extinction_Intensity} and the blue area represents the corresponding error margins. In the inset, two PLE scans of the emitter are presented to show the effect of spectral diffusion, where one is quickly scanned over the resonance (narrow line) and the other scan took several minutes. In the lower panel we show the expected behavior for the extinction measurement, when the relative phase of the local oscillator and the emitter is tuned from $\pi$ to $2 \pi$, with our measurement (dashed line) as reference.}
\end{figure*}

In a next step, we perform an extinction measurement on transition D where the waveguided excitation laser as well as the fluorescence is detected in transmission (see figure \ref{Fig:setup-overview} (a)). The principle of this measurement is based on interference in the far field between the coherent excitation field (resonant laser) and coherent photons originating from resonant fluorescence (SiV$^-$ center) \cite{Wrigge2008}. We use a $13\si{\nano\meter}$ bandpass filter centered around $740\si{\nano\meter}$ to select the coherent
single photons from the ZPL transition, as well as the coherent laser photons. We sweep the laser frequency across transition D. The resulting interference effect is presented in the top panel of fig. \ref{Fig:ResonantTransmission} (c). Here, each data point in fig. \ref{Fig:ResonantTransmission} (c) contains the averaged count rate of a 30 second count trace per frequency step. We neglect higher order modes that couple weakly to the waveguide and use the Ansatz of single mode interference between the driving field $E_{DF}$ and the resonance fluorescence $E_{RF}$ of the SiV$^-$ center. In this picture the resonance fluorescence field can be described as $E_{RF} = \alpha S(\Delta) E_{DF} \exp\left(i \varphi\right)$, where $\alpha$ and $\varphi$ are the relative weights between both fields and their relative phase \cite{Bhaskar2017}. 
The Lorentzian response of the fluorescence spectrum is given by

\begin{align}
S\left( \Delta \right) = \frac{1}{1 - \frac{2 i \Delta}{\gamma}} \, ,
\label{eq:Lorentzian_Resonance_Fluorescence}
\end{align}

where $\Delta$ describes the frequency detuning from the resonance and $\gamma$ is the FWHM of the fluorescence line shape. Therefore, the detected intensity $I_\text{det}$ is proportional to
 
\begin{align}
I_\text{det}\left( \Delta \right) \propto \left| 1 + \alpha S \left( \Delta \right) \exp \left( i \varphi \right) \right|^2 \, ,
\label{eq:Extinction_Intensity}
\end{align}
which we use to fit the data as illustrated in fig. \ref{Fig:ResonantTransmission} (c).
Due to the long measurement time, the resulting dispersive line shape contains the inhomogeneous linewidth of the emission line as seen in the inset of fig. \ref{Fig:ResonantTransmission} (c). The measurement time for the displayed PLE scans is $3.5\si{\second}$ and $100\si{\second}$ for the narrow and the broad line respectively, according to a linewidth of $\left( 154 \pm 7 \right)\si{\mega\hertz}$ and $\left( 354 \pm 9 \right)\si{\mega\hertz}$. The FWHM of the extinction curve is $\left( 360 \pm 60 \right)\si{\mega\hertz}$. 

We use excitation powers of $0.4\si{\pico\watt}$, well below the saturation powers reported for SiV centers in bulk diamond \cite{Hausler2017, Li2016}. Therefore, the condition for low power $\Omega_c/\Gamma  = 0.00249 \ll 1$ is satisfied, where $\Omega_c$ is the Rabi frequency and $\Gamma$ the total spontaneous emission rate. The relation between the relative transmitted intensity $T$ at resonance and the cooperativity $C$ in this limit can be described as $T \approx \left( C + 1 \right)^{-2}$ \cite{Chang2007}. To simulate the expected behavior of our system for different phase shifts $\varphi$, we extrapolate the relative phase in eq. \eqref{eq:Extinction_Intensity} from $\pi$ to $2 \pi$, see lower panel of fig. \ref{Fig:ResonantTransmission} (c).
By considering the relative phase of $\pi$, which represents the absorptive case, we can extract the value of the transmission intensity at resonance to $T = 0.752 \pm 0.017$. This results in a cooperativity of $C = 0.153 \pm 0.013$, which describes a lower bound for the quantum efficiency (QE) of the investigated emitter \cite{Bhaskar2017}, since the line broadening and the ZPL branching ratio \cite{Bhaskar2017, Chang2007} have an effect on the cooperativity, but not on the quantum efficiency. The lower limit for the QE determined in this way is in agreement with values from previous reports \cite{Sipahigil2016, Becker2017, Riedrich2014}. As depicted in fig. \ref{Fig:ResonantTransmission} (c), we observe line broadening beyond the lifetime limit ($154\si{\mega\hertz}/360\si{\mega\hertz} \approx 0.43$).  The beta-factor 

\begin{align}
\beta = \frac{C}{1+ C} \, ,
\label{eq:beta_factor}
\end{align}

effectively describes the coupling efficiency of the SiV$^-$ center's photon emission to the excitation laser coupled into the waveguide. Using the same approximation as for the cooperativity, the beta-factor is given by the fitting variable $\alpha$ from eq. \eqref{eq:Extinction_Intensity}, $\beta = 0.13 \pm 0.01$. The relative intensities found for the PLE measurement in transmission and reflection, compare to fig. \ref{Fig:ResonantTransmission} (b), lead to a relative coupling efficiency of $\eta_{rel} = 0.115 \pm 0.007$ and are in agreement with the results of a previous study \cite{Hoese2021}.

%Full fit function only in the supplementary material
%
%
%\begin{align}
%I \left( \Delta \right) = I_0 \left( \frac{4 \left(\Delta - \nu_0\right)^2 + \gamma^2 \left( 1 + V^2 \right) 
%+ 2 \gamma V \left( \gamma \cos\left(\varphi \right) - 2 \left(\Delta - \nu_0\right) \sin\left( \varphi \right) \right)}{4 \left(\Delta - \nu_0\right)^2 + \gamma^2} \right)
%\label{eq:Extinction_Intensity_Fit}
%\end{align}

\section{Super-Poissonian light statistics by single photon absorption}

\begin{figure}[]
\includegraphics[scale=1.0]{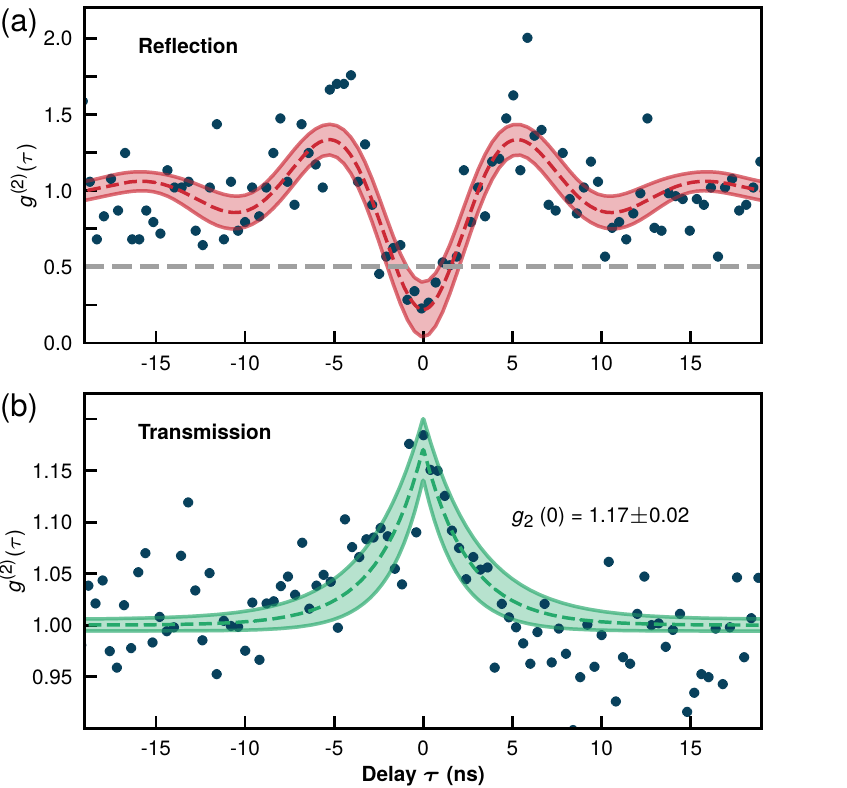}
\caption{\label{Fig:resonant-correlation}Second order correlation measurements in reflection and transmission. 
(a) Results of a second order correlation measurement in reflection at resonance ($739.338\si{\nano\meter}$) with a SiV$^-$ center ZPL transition, where only PSB photons are detected. Here, a fitted value of $g^{(2)}(0)<0.5$ corresponds to single photon emission by the defect center. 
(b) Correlation measurement at resonance with the same transition, detected in transmission ($740/13\si{\nano\meter}$ BP). Here, the quasi-coherent laser light is correlated with the emitters fluorescence, which is guided through the waveguide (compare to fig. \ref{Fig:setup-overview} (b)). At zero time delay bunching is clearly visible. The determined error margins are indicated by the highlighted red and green areas respectively.} 
\end{figure}

Now we turn to the quantum non-linear character of the SiV$^-$-waveguide system. Therefore, we measure the autocorrelation of the transmitted signal in a HBT configuration. First, we measure the autocorrelation in reflection by exciting the defect center at resonance and detecting only the PSB (see fig. \ref{Fig:resonant-correlation} (a)). Since $g^{\left(2\right)}\left(0\right)$ is well below the threshold of 0.5 for a single photon emitter, we conclude that only a single defect center is excited. The transmission signal, obtained under identical conditions as the extinction measurement but for a quasi-coherent field in fig. \ref{Fig:ResonantTransmission} (c), should now be modified by the non-linearity of the single photon transition of the SiV$^-$ center. In order to investigate the non-linearity, we work with sufficiently low laser powers, so that the driving field is a quasi-coherent state with non-negligible single- and two-photon contribution. Within the fluorescence lifetime of the emitter, the quasi-coherent field has a mean photon number of $0.00223$, which is determined with a power meter. Hence, the quasi-coherent state, normalized to the emitter's lifetime, is described by

\begin{align*}
\ket{\alpha} & =  e^{-\frac{\left| \alpha \right|^2}{2} } \sum_{n=0}^{\infty} \frac{\alpha^n}{\sqrt{n!}} \ket{n} \\& =  0.9989 \ket{0} + 0.0472 \ket{1} +  0.0016 \ket{2} +... \, .
\end{align*}
%=  0.99889 \ket{0} + 0.04719 \ket{1} +  0.00158 \ket{2}+0.00004 \ket{3}
In the case of large photon-atom interaction a low photon flux of the excitation field can saturate the defect center's transition and a single SiV$^-$ center can modify the output field. Consequently, single photons are removed from the quasi-coherent state resulting in a super-Poissonian light statistic of the transmitted light. In this case a bunching behavior at zero time delay is expected in the HBT  correlation measurement, as confirmed by the autocorrelation measurement presented in fig. \ref{Fig:resonant-correlation} (b). We use
\begin{align}
g^{\left( 2\right)} \left( \tau \right) = 1 - a e^{-3 \Gamma /4 \left| \tau \right|}\left( \cos \Omega_\Gamma \left| \tau \right|  + \frac{3 \Gamma}{4 \Omega_\Gamma} \sin \Omega_\Gamma \left| \tau \right|   \right) \, ,
\label{eq:g2_2lvl_res}
\end{align} 
to fit the antibunching curve, since we excite the transition with a strong driving field \cite{Steck2007}, where $\Gamma$ is the emitters fluorescence decay rate, $\Omega_\Gamma = (\Omega_c^2 - \left(\Gamma/4\right)^2)^{0.5}$ is the damped Rabi frequency and $\Omega_c$ is the Rabi frequency. To evaluate the bunching behavior in the transmission measurement, we use the fit function
\begin{align}
g^{\left( 2\right)} \left( \tau \right) = 1 + a e^{-\Gamma \left| \tau \right|} \, .
\label{eq:g2_2lvl}
\end{align}
%$g^{\left( 2\right)} \left( 0 \right) = 1.40 \pm 0.08$
At zero time delay we get $g^{\left( 2\right)} \left( 0 \right) = 1.17 \pm 0.02$. The bunching behavior is confirmed by two other measurements as illustrated in the supplementary information. As control we measure the correlation function for the laser (far detuned from resonance) which results in a flat line, with $g^{\left( 2\right)} \left( \tau \right) = 1 $ as expected for a coherent state (see supplementary information).  
With that we can confirm that the observed super-Poissonian light statistic of the investigated defect center is the result of single photon interference in the far field (detector location)\cite{Bhaskar2017, Chang2007, Javadi2015}.

\section{Discussion and Outlook}
In summary, by combining high-NA excitation with waveguide-assisted detection we achieve a high cooperativty for laser-written diamond photonics despite its large mode-field diameter. We present evidence for a single photon non-linearity in a laser-written SiV$^-$-waveguide system, where the non-linearity arises from the interaction between the quasi-coherent excitation field and the resonant fluorescence of the defect center. 
We perform an extinction measurement on the SiV$^-$-waveguide system in order to extract the cooperativity of $15.3\si{\percent}$ yielding the lower bound of a single emitter's QE. Furthermore, we probe the polarization-selectivity of the system. Ultimately, we demonstrate super-Poissonian light statistics originating from single photon subtraction of a quasi-coherent light field. This opens a new path to the longstanding goal to engineer light at the single photon level. 

In the future, controlling the phase between excitation and emission path will further enable switching between single photon subtraction and addition leading to the engineering of arbitrary, nonclassical light fields \cite{Du20}. 
The non-classical light field engineering could build on the ability to laser-write advanced photonic components such as Y-beamsplitter and Bragg-gratings \cite{Courvoisier2016, Bharadwaj2017} and benefits from the three-dimensional writing capabilities. Functionalization of multiple waveguides by wide-field implantation of the front facet enables scalable quantum technology and good optical access via standard optical fibers.
Extending Y-beamsplitters to X-beamsplitter facilitates, together with the high degree of indistinguishability of single photons emitted from SiV$^-$ centers, on-chip Hong-Ou-Mandel interference \cite{Sipahigil2014} and the construction of path-entangled light fields for applications in quantum metrology, such as super-resolution phase estimation beyond the standard quantum limit \cite{Polino2020}.

\section*{Acknowledgments}
The authors would like to thank Felix Breuning for experimental support in the beginning of the project. IFN-CNR and UUlm are grateful for support from the H2020 Marie Curie ITN project LasIonDef (GA n.956387). IFN-CNR is thankful for support from the project QuantDia (FISR2019-05178) funded by Ministero dell’Istruzione, dell’Universit\`a e della Ricerca.
A.K. acknowledges support of the European Regional Development Fund (EFRE) program Baden-Württemberg. M.K.K. and A.K. acknowledge support of IQst. M.H. acknowledges support from the Studienstiftung des deutschen Volkes. V.B acknowledges the support of the Alexander von Humboldt Foundation

\section*{Author Contributions}
M.K.K., M.H., V.B., S.M.E. and A.K. conceived the project. M.K.K. and M.H. performed and analyzed all PL, PLE and correlation measurements. V.B., J.P.H., R.R. and S.M.E. produced the laser-written WG in diamond, which were implanted and annealed by J.L. and F.J. to create SiV$^-$ centers. All authors discussed the results. M.K.K., M.H. and A.K. wrote the manuscript, which was discussed and edited by all authors.

% Create the reference section using BibTeX:
\bibliography{WG-SiVTransmission}

\end{document}